\documentclass[twocolumn]{aastex701}
\usepackage{natbib}
\usepackage{amsmath}
\usepackage{graphicx,caption}

\shorttitle{How Eccentric are Planet-Perturbed Disks?}
\shortauthors{Padgett et al.}
\graphicspath{{./}{Figures/}}

\begin{document}
%%%%%%%%%%%%%%%%%%%%%%%%%%%%%%%%%%%%%%%%%%%%%%%%%%%%%%%%%%%%%%%%%%%%%%%%%%%%%%%%%%%%%%%%%%%
\title{On Eccentric Protoplanetary Disks I --- How Eccentric are Planet-Perturbed Disks?}

%%%%%%%%%%%%%%%%%%%%%%%%%%%%%%%%%%%%%%%%%%%%%%%%%%%%%%%%%%%%%%%%%%%%%%%%%%%%%%%%%%%%%%%%%%%
%%%%%%%%%%%%%%%%%%%%%%%%%%%%%%%%%%%%%%%%%%%%%%%%%%%%%%%%%%%%%%%%%%%%%%%%%%%%%%%%%%%%%%%%%%%
%%%%%%%%%%%%%%%%%%%%%%%%%%%%%%%%%%%%%%%%%%%%%%%%%%%%%%%%%%%%%%%%%%%%%%%%%%%%%%%%%%%%%%%%%%%
\author[0009-0006-6220-1811]{Cory Padgett}
\affiliation{Department of Physics and Astronomy,\\ 118 Kinard Laboratory, Clemson University,\\ Clemson, SC 29634-0978, USA}
\email[show]{cpadge4@clemson.edu}  

\author[0000-0002-7455-6242]{Jeffrey Fung}
\affiliation{Department of Physics and Astronomy,\\ 118 Kinard Laboratory, Clemson University,\\ Clemson, SC 29634-0978, USA}
\email{fung@clemson.edu}  

%%%%%%%%%%%%%%%%%%%%%%%%%%%%%%%%%%%%%%%%%%%%%%%%%%%%%%%%%%%%%%%%%%%%%%%%%%%%%%%%%%%%%%%%%%%
%%%%%%%%%%%%%%%%%%%%%%%%%%%%%%%%%%%%%%%%%%%%%%%%%%%%%%%%%%%%%%%%%%%%%%%%%%%%%%%%%%%%%%%%%%%
%%%%%%%%%%%%%%%%%%%%%%%%%%%%%%%%%%%%%%%%%%%%%%%%%%%%%%%%%%%%%%%%%%%%%%%%%%%%%%%%%%%%%%%%%%%
\begin{abstract}
Protoplanetary disks can become eccentric when planets open deep gaps within, but how eccentric are they? We answer this question by analyzing two-dimensional hydrodynamical simulations of planet-disk interaction. The steady state eccentricity of the outer disk (outside of the planet's orbit) is described as a balance between eccentricity excitation by the 1:3 eccentric Lindblad resonance and eccentricity damping by gas pressure. This eccentricity scales with $q(h_{\rm p}/r_{\rm p})^{-1}\left({r_{\rm gap}}/{r_{\rm p}}\right)^{a-\frac{b}{2}-2}$, where $q$ is the planet-to-star mass ratio, $h_{\rm p}/r_{\rm p}$ is the disk aspect ratio, $r_{\rm gap}/r_{\rm p}$ is the radial position of the outer gap edge divided by the planet's position, and $a$ and $b$ are the negative exponents in the disk's surface density and temperature power law profiles, respectively. We derive a semi-analytic eccentricity profile that agrees with numerical simulations to within $30\%$. Our result is a first step to quantitatively interpret observations of eccentric protoplanetary disks, such as MWC 758, HD 142527, IRS 48, and CI Tau.
\end{abstract}

%%%%%%%%%%%%%%%%%%%%%%%%%%%%%%%%%%%%%%%%%%%%%%%%%%%%%%%%%%%%%%%%%%%%%%%%%%%%%%%%%%%%%%%%%%%
%%%%%%%%%%%%%%%%%%%%%%%%%%%%%%%%%%%%%%%%%%%%%%%%%%%%%%%%%%%%%%%%%%%%%%%%%%%%%%%%%%%%%%%%%%%
%%%%%%%%%%%%%%%%%%%%%%%%%%%%%%%%%%%%%%%%%%%%%%%%%%%%%%%%%%%%%%%%%%%%%%%%%%%%%%%%%%%%%%%%%%%
%Unified Astronomy Thesaurus concepts

%%%%%%%%%%%%%%%%%%%%%%%%%%%%%%%%%%%%%%%%%%%%%%%%%%%%%%%%%%%%%%%%%%%%%%%%%%%%%%%%%%%%%%%%%%%
%%%%%%%%%%%%%%%%%%%%%%%%%%%%%%%%%%%%%%%%%%%%%%%%%%%%%%%%%%%%%%%%%%%%%%%%%%%%%%%%%%%%%%%%%%%
%%%%%%%%%%%%%%%%%%%%%%%%%%%%%%%%%%%%%%%%%%%%%%%%%%%%%%%%%%%%%%%%%%%%%%%%%%%%%%%%%%%%%%%%%%%
\section{Introduction} \label{sec:Intro}
Planets embedded in their natal protoplanetary disks can excite disk eccentricities. Indeed, eccentric protoplanetary disks have been observed, including MWC 758 \citep{Dong2018,Kuo2022}, HD 142527 \citep{Garg2021}, IRS 48 \citep{Yang2023}, and CI Tau \citep{Kozdon2023}. However, quantitatively linking these eccentric disks to their potential planetary origins has been challenging, mainly because we lack a predictive model for the fully nonlinear state of planet-excited disk eccentricity.

Planets affect disk eccentricity through resonance interactions. Pioneering work \citep{Goldreich1981,Hirose1990,Lubow1991} demonstrated that the presence of a planet can amplify disk eccentricity at eccentric Lindblad resonances (ELR; specifically, the resonances that are not co-rotating with the planet), but eccentric co-rotational resonances (ECR), resonances that co-rotate, and therefore coincide, with the regular Lindblad resonances, damp disk eccentricity. Generally, we expect disks to be nearly circular because ELRs and ECRs overlap with each other, and ECR are typically stronger. However, the 1:3 ELR, which lies furthest away from the planet, does not overlap with any ECR. When a planet opens a wide, deep gap, it can deplete nearly all ELR and ECR, leaving the 1:3 ELR as the last surviving resonance. In this case, it is expected that disk eccentricity will grow exponentially. This linear growth has been studied extensively \citep{Teyssandier2016,Teyssandier2019}, and has also been shown in numerical simulations \citep[e.g.,][]{Papaloizou2001,Kley2006,Tanaka2022}.

The eccentricity, $e$, of a disk reaches steady state when the 1:3 ELR is countered by some opposing effect. Previous work have largely focused on the damping effects of bulk viscosity \citep[e.g.,][]{Ogilive2001,Goodchild2006}, which is likely because gas pressure does not damp or excite eccentricity when considering a linear (i.e,  of order $e$) perturbation \citep{Lubow1991}. However, in the fully nonlinear case, gas pressure is still likely a damping force. After all, eccentricity introduces velocity variations that compress and expand the gas, which gas pressure naturally acts against. Since its linear term, expanded in $e$, does not damp eccentricity, perhaps the next largest term, of order $e^2$, should be considered. In other words, the damping force due to gas pressure might be of order $e^2(h/r)^2$, where $h/r$ is the disk's aspect ratio and its square describes the magnitude of gas pressure. Meanwhile, the strength of the 1:3 ELR is proportional to $q^2$ \citep[][]{Lubow1991,Lubow2010}, where $q$ is the planet-to-star mass ratio. By balancing the excitation and damping terms, we might expect the proportionality: $q^2\propto e^2(h/r)^2$, for some measurement of the ``total'' eccentricity of the disk. Or, 
\begin{equation}
    e \propto \frac{q}{(h/r)} \, .
    \label{eq:scale}
\end{equation}

Without knowing how $e$ varies spatially, we expect this scaling to hold in a zeroth-dimensional, holistic sense. The key to uncovering this scaling is to properly define the disk's ``total eccentricity''. To that end, it is necessary to clarify the meaning of eccentricity, in the context of a gaseous disk that is strongly affected by nonlinear perturbations.

Gaseous disk do not follow Keplerian orbits exactly. In a significantly eccentric disk, gas will experience time-dependent torques due to varying pressure gradients along its orbit. Notably, these gradients do not need to be small. In fact, they need to be strong enough to counter eccentricity excitation, which is itself strong enough to make the disk significantly eccentric. In other words, the star, the planet, and gas pressure can all be exerting order-unity influence on the orbit's shape, and there is no clear reason to assume the resultant shape is close to Keplerian (i.e., elliptical). While Keplerian solutions can still serve as analytic guides to our understanding, direct applications of those solutions, such as the definition of eccentricity, may lead to confusing interpretations.

We have therefore opted to measure ``eccentricity'' more literally as a measurement of the disk's off-centered-ness, or, in other words, its $m=1$ mode, where $m$ is the usual azimuthal mode number. Furthermore, we restrict ourselves to the $m=1$ mode of the disk's radial velocity, $v_{r}$, to ensure that we are truly detecting the part of the disk that has an apoapsis and a periapsis, rather than disk structures, like vortices, that would contribute to the $m=1$ signal in disk density even when orbits are nearly circular. Our definition of eccentricity is expressed as:
\begin{equation}
    e_1 (r) \equiv \frac{2\left|\int v_r(r,\theta) ~ e^{i\theta} ~{\rm d}\theta\right|}{\int r\Omega(r,\theta) ~{\rm d}\theta} \, ,
    \label{eq:e1}
\end{equation}
where $\Omega$ is the disk's angular speed, and $r$ and $\theta$ are the usual polar coordinates centered at the star. The subscript $1$ is a reminder that our ``eccentricity'' is in truth the $m=1$ mode amplitude. The factor of $2$ in the numerator ensures that our expression is equal to the two-body problem definition of eccentricity, in the $e\rightarrow0$ limit; i.e., when $v_{r} \approx e v_{\rm k} \sin(\theta-\theta_0)$, where $v_{\rm k}$ is the Keplerian speed and $\theta_0$ is the argument of periapsis. For the rest of this paper, we will refer to $e_1$ as just eccentricity, but hope the reader keeps the subtle distinction in mind.

For our measurement of ``total'' eccentricity, we multiply $e_1$ by the disk's angular momentum density and take the azimuthal average:
\begin{equation}
    \ell_1 (r) \equiv \frac{1}{2\pi}\int_0^{2\pi} e_1 \Omega\Sigma r^3 ~{\rm d}\theta \, ,
    \label{eq:l1}
\end{equation}
where $\Sigma$ is the disk surface density. Then, we integrate it radially outward of the planet's position to get a measurement for the outer disk:
\begin{equation}
    \mathcal{L}_1 \equiv \int_{r_{\rm p}}^{r_{\rm out}} \ell_1 ~{\rm d}r \, ,
    \label{eq:L1}
\end{equation}
where $r_{\rm p}$ is the radial position of the planet and $r_{\rm out}$ is the outer edge of the disk. A similar measurement can be made for the inner disk as well. This definition of $\mathcal{L}_1$ can be associated with the concept of total angular momentum deficit that has been used in the past \citep[e.g.,][]{Ragusa2018,Ragusa2020}. This may seem surprising given that the angular momentum deficit for Keplerian orbits scales with $e^2$, whereas our $\mathcal{L}_1$ scales linearly with $e_1$. Here, we again try to steer away from the idea of Keplerian orbits. The $e^2$ scaling arises from the fact that angular momentum deficit scales with $1-\sqrt{1+e\cos{\theta}}$, and the linear term in this factor vanishes over azimuthal averaging, but what if $e$ is not azimuthally constant? If $e$ also has some periodic azimuthal variations (i.e., orbits are not Keplerian), then, generally, the linear term will not vanish. As we will see later, $\mathcal{L}_1$ does turn out to be an exceptionally good diagnostic for describing the fully non-linear steady state of a planet-perturbed disk.

%%%%%%%%%%%%%%%%%%%%%%%%%%%%%%%%%%%%%%%%%%%%%%%%%%%%%%%%%%%%%%%%%%%%%%%%%%%%%%%%%%%%%%%%%%%
%%%%%%%%%%%%%%%%%%%%%%%%%%%%%%%%%%%%%%%%%%%%%%%%%%%%%%%%%%%%%%%%%%%%%%%%%%%%%%%%%%%%%%%%%%%
%%%%%%%%%%%%%%%%%%%%%%%%%%%%%%%%%%%%%%%%%%%%%%%%%%%%%%%%%%%%%%%%%%%%%%%%%%%%%%%%%%%%%%%%%%%
\section{Methods} \label{sec:Methods}
\begin{figure*}[]
    \centering
    \includegraphics[width=0.95\linewidth]{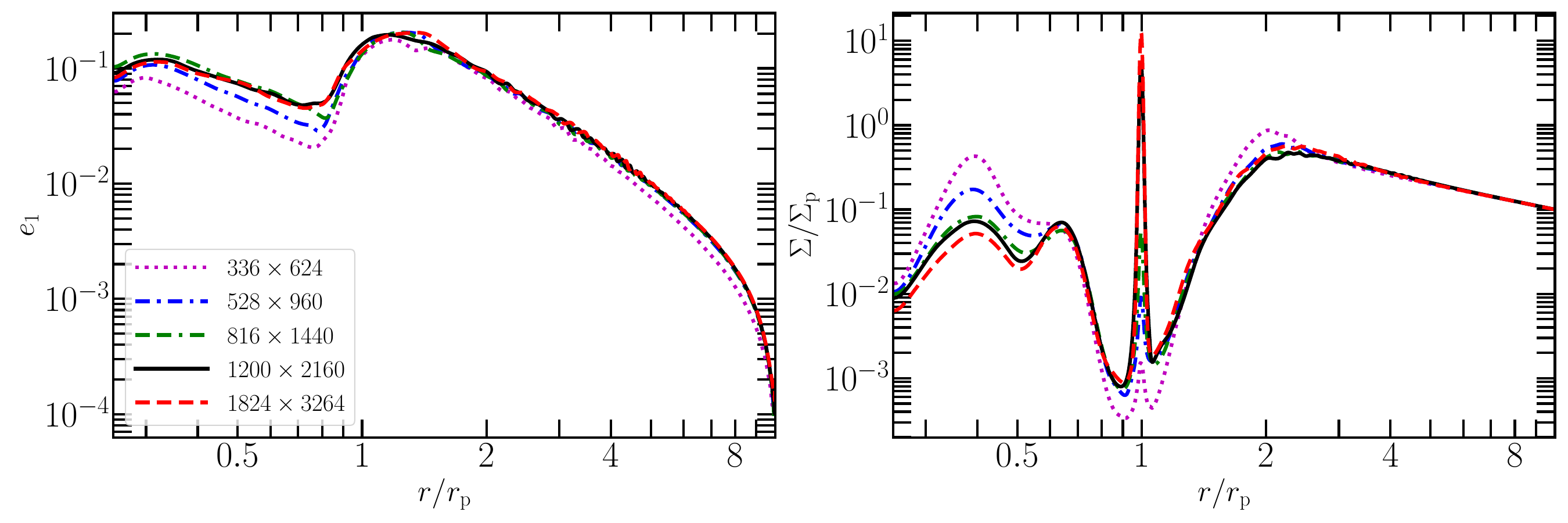}
    \caption{Azimuthally averaged eccentricity (left) and surface density (right) at different resolutions. The parameters for these simulations are $\{q,h/r,a,b\}=\{0.004,0.05,1,1\}$. Our simulations converge well at our choice of resolution, $1200 (r)\times2160(\theta)$, especially in the outer disk ($r>r_{\rm p}$) where we focus our analysis.}
    \label{fig:convergence}
\end{figure*}

%%%%%%%%%%%%%%%%%%%%%%%%%%%%%%%%%%%%%%%%%%%%%%%%%%%%%%%%%%%%%%%%%%%%%%%%%%%%%%%%%%%%%%%%%%%
\subsection{Numerical setup} \label{subsec:NUmSet}
We use the graphics processing unit-accelerated hydrodynamics code \texttt{PEnGUIn} \citep{Fung_thesis} to simulate two-dimensional (2D), protoplanetary disks with embedded planets in polar ($r$-$\theta$) coordinates. \texttt{PEnGUIn} solves the Navier-Stokes equations:
\begin{equation} \label{continuity_eq}
    \frac{D\Sigma}{Dt} = -\Sigma (\nabla \cdot \textbf{v}) \, ,
\end{equation}
\begin{equation} \label{momentum_eq}
    \frac{D \textbf{v}}{Dt} = - \frac{1}{\Sigma}\nabla P - \nabla \Phi + \frac{1}{\Sigma} \nabla\cdot \mathbf{T} \, ,
\end{equation}
\begin{equation} \label{energy_eq}
    \frac{D u}{Dt} = -\frac{P}{\Sigma} \nabla\cdot \textbf{v} -\Lambda \, ,
\end{equation}
where $\Sigma$ is the gas surface density, $P$ is the vertically-integrated gas pressure, $\Phi$ is the gravitational potential of the star and planet, $\textbf{v}={v_r}\hat{r} + r\Omega\hat{\theta}$ is the gas velocity, $v_r$ is the radial velocity, $\Omega$ is the angular velocity, $u$ is the specific internal energy, and $\Lambda$ is the cooling rate. $\mathbf{T}$ is the Newtonian stress tensor containing both the bulk and shear viscosity components:
\begin{align}
    \mathrm{T}_{r,r} &= 2\nu\Sigma \frac{\partial v_{r}}{\partial r} \, , \\
    \mathrm{T}_{\theta,\theta} &= 2\nu\Sigma \left(\frac{1}{r}\frac{\partial r\Omega}{\partial \theta} +  \frac{v_r}{r}\right) \, , \\
    \mathrm{T}_{r,\theta} &= \nu\Sigma \left(\frac{1}{r}\frac{\partial v_{r}}{\partial \theta} + r \frac{\partial \Omega}{\partial r}\right) \, , 
\end{align}
where $\nu$ is the kinematic viscosity. We use the Sunyaev-Shakura $\alpha$-viscosity prescription \citep{alpha-viscosity} and set $\nu = \alpha c_{\rm s} h$, where $c_{\rm s}$ is the gas sound speed and $h$ is the local disk scale height. Our expression for $\mathbf{T}$ is equivalent to setting the bulk viscosity to be $5/3$ times the shear viscosity. This choice ensures that our disks are stable to eccentricity excitation driven by shear viscosity \citep[see][]{Lyubarskij1994,Ogilive2001,Latter2006}.

We adopt a polytropic equation of state:
\begin{equation}
    u=\frac{P}{(\gamma-1)\Sigma}\, , 
\end{equation}
where $\gamma$ is the ratio of specific heats, and is set to $7/5$ for all of our runs. The exact value of $\gamma$, however, is inconsequential because the cooling rate, $\Lambda$ in eq. \ref{energy_eq}, follows a simple Newtonian prescription:
\begin{equation}
    \Lambda = \frac{u-u_0}{t_{\rm cool}}
\end{equation}
where the relaxation time $t_{cool}$ is set to be $10^{-2}\Omega_{\rm K}^{-1}$ for all runs. We have tested and verified that, given this $t_{\rm cool}$, our disks are effectively locally isothermal at all times across all models. This is to simplify our analysis and reduce the degrees of freedom in our models. We will discuss the potential effects of choosing longer $t_{\rm cool}$ in sec. \ref{sec:Discussion}. The equilibrium value $u_0$ is a prescribed specific internal energy profile determined by the initial conditions.

The star and planet gravitational potential in inertial frame can be expressed as:
\begin{equation}
    \begin{split}
        \Phi =& -\frac{GM_*}{\sqrt{r^2+r_1^2+2rr_1 \cos(\theta-\theta_{\rm p})}} \\
        &-\frac{GM_{\rm p}}{\sqrt{r^2+r_2^2-2rr_2 \cos(\theta-\theta_{\rm p})}+r_{\rm s}^2}  \, ,
    \end{split}
    \label{eq:potential}
\end{equation}
and $q =  M_{\rm p} /M_*$ is the planet-to-star mass ratio, where $M_*$ and $M_{\rm p}$ are the masses of the star and the planet, respectively. The star's and planet's radial positions are $r_1 = q r_{\rm p} / (1+q)$ and $r_2 = r_{\rm p} / (1+q)$ and their angular positions are $\theta_{\rm p}-\pi$ and $\theta_{\rm p}$. We adopt a smoothing length, $r_{\rm s}$, to soften the planet's gravitational potential and to mimic the torque density of a three-dimensional simulation. We use $r_{\rm s}=0.7h$ for all runs which is appropriate for gap-opening simulations \citep[][]{Muller2012,Fung2016}. The planet's mass is gradually inserted over a number of orbits to avoid strong initial shocks; which we set as $10$ orbits for all runs. Additionally, we implemented orbital advection, rotating the grid in the frame of the orbiting planet to reduce the computational load.

We choose a grid of $1200 \times 2160$ radial-azimuthal cells for a space of $0.25 r_{\rm p}$ to $10 r_{\rm p}$ as our nominal resolution. Fig. \ref{fig:convergence} demonstrates that, with this resolution, our disk profiles are numerically converged.

%%%%%%%%%%%%%%%%%%%%%%%%%%%%%%%%%%%%%%%%%%%%%%%%%%%%%%%%%%%%%%%%%%%%%%%%%%%%%%%%%%%%%%%%%%%
\subsubsection{Initial and Boundary Conditions}
\label{sec:Initial_Bound}
The initial disk is in hydrostatic equilibrium with respect to the star's gravity. The surface density and temperature profiles are:
\begin{equation}\label{disk_dens_struct}
    \Sigma = \Sigma_{\rm p} \left( \frac{r}{r_{\rm p}} \right)^{-a}  \, ,
\end{equation}
\begin{equation}\label{disk_temp_struct}
    c_{\rm s}^2 = c_{\rm s, p}^2 \left( \frac{r}{r_{\rm p}} \right)^{-b}  \, ,
\end{equation}
where $\Sigma_{\rm p}$ and $c_{\rm s,p}^2$ are the initial unperturbed background normalizations of the density and temperature profiles at the planet's orbital radius. The initial gas pressure is then simply $P=c_{\rm s}^2\Sigma$. The value of $\Sigma_{r_{\rm p}}$ is inconsequential since we do not calculate disk self-gravity (i.e., Equation \ref{momentum_eq}). The value of $c_{\rm s,r_{\rm p}}^2$ is set according to the model's value of $h/r$, the disk aspect ratio, which ranges between $0.035$ and $0.1$ at the planet's location. The profile of $\alpha$ is chosen to ensure that our initial condition has a steady viscous background flow. This requires that $\nu\Sigma$ be a constant, which, translating to $\alpha$, is equivalent to:
\begin{equation}\label{disk_vis_struct}
    \alpha = \alpha_{\rm p} \left( \frac{r}{r_{\rm p}} \right)^{a+b-\frac{3}{2}} \, .
\end{equation}
If $\alpha$ follows any other profile, we expect $\Sigma$ to evolve away from its initial condition in order to maintain a steady viscous flow. With our choice, we find that the outer disk largely remains close to its initial density, allowing us to approximate the steady-state profile with the initial profile. Since bulk viscosity also can damp eccentricity driven by the planet, we isolate its effects by choosing $\alpha_{\rm p}$ values that are sufficiently small so that eccentricity damping is dominated by gas pressure. We typically find that this requires $\alpha_{\rm p}$ to be $10^{-3}$ or smaller. 

The azimuthal boundary condition is periodic and spans the full $2\pi$ range where the cells are linearly spaced. The inner radial boundary employs a zero-gradient, outflow boundary condition. The choice of using an outflow boundary at the inner edge is atypical. More commonly, planet-disk interaction simulations employ fixed boundaries at both the inner and outer bounds. However, we have experimented and verified that a fixed inner boundary tends to artificially circularize the inner disk. \cite{Kley2008} similarly found the choice of inner boundary condition strongly affects the inner disk's evolution. While our choice frees the inner disk from artificial eccentricity damping, it may lead to other numerical artifacts. We echo \cite{Kley2008} and recognize that there is a general need to better understand the effects of different boundary conditions in planet-disk interaction simulations.

The evolution of the inner disk is additionally complicated by the fact that it is fed by material that has to pass through the planet's orbit; meanwhile, the outer disk is viscously fed by the outer boundary, which we can control in our simulations. For these reasons, our analysis will focus exclusively on the outer disk. We defer a description of the inner disk's evolution to the second paper of this series.

The radial cells are logarithmically spaced and have an inner radius of $0.25 r_{\rm p}$ and an outer radius of $10 r_{\rm p}$. The outer boundary is fixed for all fluid variables as described by our initial conditions. This ensures that in steady-state, the disk must be in equilibrium with the prescribed amount of inflow. However, this also means the outer boundary is fixed to be axisymmetric, i.e., circular. This creates significant artificial eccentricity damping near the outer boundary, as shown in fig. \ref{fig:convergence} and throughout sec. \ref{sec:Results}. By placing the outer boundary far from the planet, at $10 r_{\rm p}$, we have attempted to minimize this effect. The commonly employed wave-killing zone \citep{DeValBorro2006} is not used because we observe no significant wave reflections at the outer boundary. Density perturbations typically weaken to below $1\%$ before reaching the outer boundary.

%%%%%%%%%%%%%%%%%%%%%%%%%%%%%%%%%%%%%%%%%%%%%%%%%%%%%%%%%%%%%%%%%%%%%%%%%%%%%%%%%%%%%%%%%%%
%%%%%%%%%%%%%%%%%%%%%%%%%%%%%%%%%%%%%%%%%%%%%%%%%%%%%%%%%%%%%%%%%%%%%%%%%%%%%%%%%%%%%%%%%%%
%%%%%%%%%%%%%%%%%%%%%%%%%%%%%%%%%%%%%%%%%%%%%%%%%%%%%%%%%%%%%%%%%%%%%%%%%%%%%%%%%%%%%%%%%%%
\section{Results} \label{sec:Results}
\begin{figure*}[!p]
    \centering
    \includegraphics[width=0.75\linewidth]{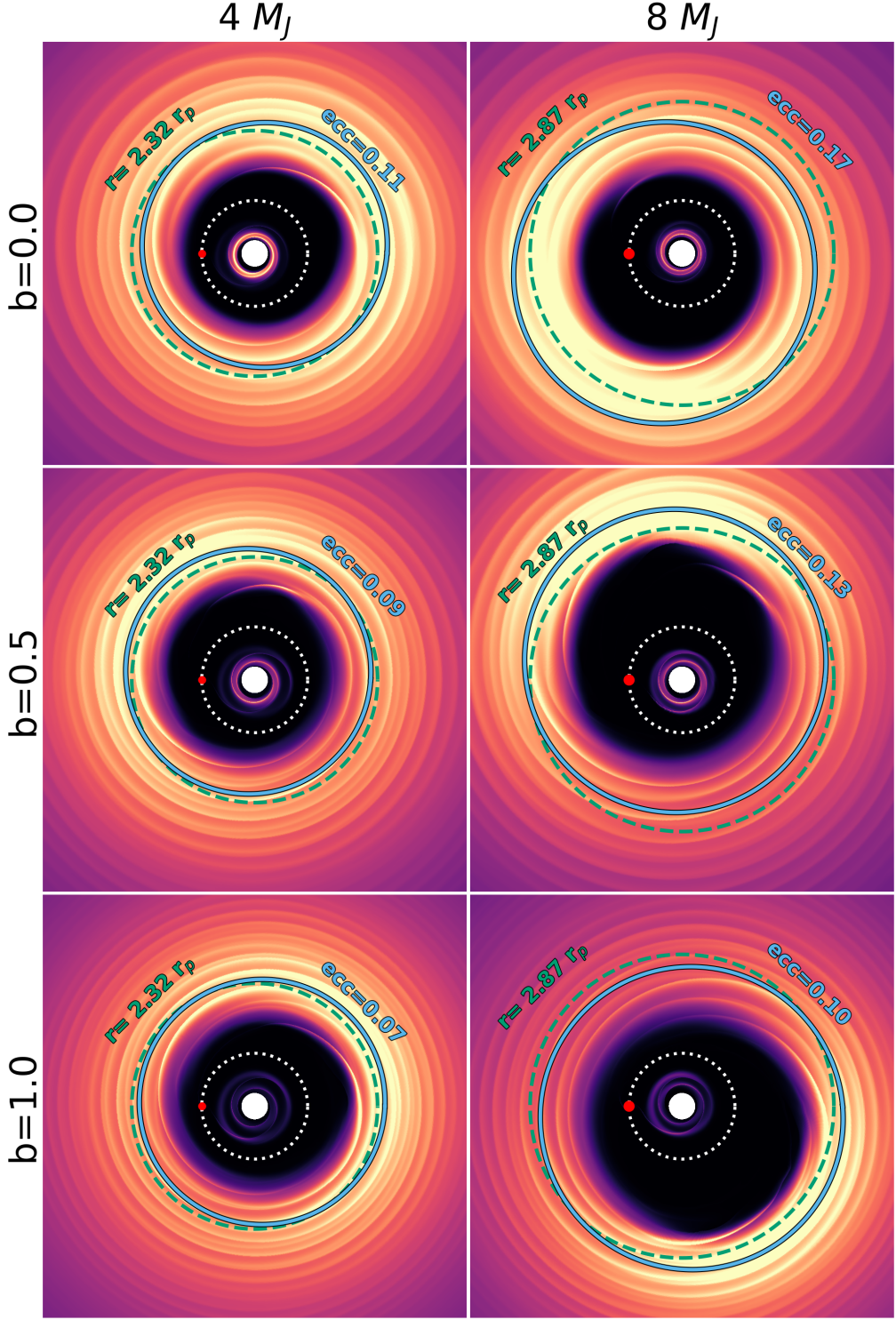}
    \caption{A gallery of our simulations displaying how disk morphology changes with the planet-to-star mass ratio $q$ and the surface temperature profile's (eq. \ref{disk_temp_struct}) parameter $b$. Other parameters are fixed at $\{h_{\rm p}/r_{\rm p}, a, \alpha_{\rm p}\}=\{0.05, 1, 10^{-3}\}$. The image color gradient is linear scale from $0.0-0.5 \hspace{0.1cm} \Sigma_p$. Planets' locations and orbits are de-marked with a red dot and a white dotted line. Green dashed circles references the locations of $r_{\rm gap}$ (eq. \ref{eq:rgap}), the outer gap edges if they were circular. The true gap edges are eccentric, and are represented by the blue ellipses that have semi-major axis $r_{\rm gap}$ and eccentricity $e_{\rm gap}$ (eq. \ref{eq:e_gap}).}
    \label{fig:disks}
\end{figure*}

We performed a suit of hydrodynamical simulations with different combinations of the values $q$, $h_{\rm p}/r_{\rm p}$, $a$, and $b$. The values of $q$ are $0.004$ and $0.008$; the values of $h_{\rm p}/r_{\rm p}$ are $0.1$, $0.05$, and $0.035$; $a$ is varied between $0.5$ and $2.0$, and $b$ is between $0$ and $1$. The value of $\alpha_{\rm p}$ should be inconsequential as long as it small. We typically choose $10^{-3}$, except when $h_{\rm p}/r_{\rm p}=0.1$, then $\alpha_{\rm p}$ is lowered to $10^{-4}$. Across this parameter space, the measured steady-state eccentricity at the outer gap edge typically lies at $\sim 0.05-0.14$ (see sec. \ref{sec:wide_gap} for exceptions). Fig. \ref{fig:disks} is a gallery of a subset of our simulations.

Simulations typically last $10^4$ planetary orbits, but those with $\alpha_{\rm p}=10^{-4}$ are extended to $2\times 10^{4}$ orbits to account for their longer viscous times. All profiles presented are temporal averages over the last $10^3$ orbits. In our parameter space, planets typically open a deep gap and begin exciting disk eccentricity within the first $10^3$ orbits. As discussed in the previous section, we will focus our analysis on the outer disk (i.e., outward of the planet's orbit) only.

We measure the quantity $\mathcal{L}_1$ (eq. \ref{eq:L1}) and inspect whether it follows our expected scaling (eq. \ref{eq:scale}). For this, we extend the concept behind eq. \ref{eq:scale} and argue that eccentricity excitation should overall scale with both $q$ and the surface density at the 1:3 ELR, which is located at $r_{1:3}=3^{\frac{2}{3}}r_{\rm p}\approx2.08r_{\rm p}$. Meanwhile, eccentricity damping might scale with the magnitude of gas pressure evaluated at the most eccentric part of the disk, which is likely where it is closest to the planet, i.e., the outer edge of the planetary gap. We denote this radial location as $r_{\rm gap}$. The magnitude of gas pressure is proportional to $(h/r)^2$ times the surface density, both evaluated at $r_{\rm gap}$.  Balancing these two quantities gives us a scaling factor similar to eq. \ref{eq:scale}, but with an additional factor that accounts for the different gas densities at $r_{1:3}$ and $r_{\rm gap}$. We name this quantity $Q$:
\begin{equation}
    Q \equiv q\left(\frac{h(r_{\rm gap})}{r_{\rm gap}}\right)^{-1}\left(\frac{r_{1:3}}{r_{\rm gap}}\right)^{-a} \, .
    \label{eq:Q}
\end{equation}
Because the temperature profile exponent $b$ varies across different simulations, even models with the same $h_{\rm p}/r_{\rm p}$ value can have different values of $h/r$ at $r=r_{\rm gap}$. \cite{Kanagawa2016} suggested that for deep gaps, the gap width can be predicted by the formula:
\begin{equation}
    \frac{r_{\rm gap}}{r_{\rm p}} = 1+0.41 q^{\frac{1}{2}}\left( \frac{h_{\rm p}}{r_{\rm p}}\right)^{-\frac{3}{4}}\alpha_{\rm p}^{-\frac{1}{4}} \, ,
    \label{eq:rgap}
\end{equation}
which we will adopt for our analysis. It is common to define:
\begin{equation}
    K' \equiv q^{2}\left( \frac{h_{\rm p}}{r_{\rm p}}\right)^{-3}\alpha_{\rm p}^{-1} \, ,
    \label{eq:K_prime}
\end{equation}
so that $r_{\rm gap}/r_{\rm p}$ can also be written more concisely as $1+0.41{K'}^{\frac{1}{4}}$. We note that the definition of gap edge is somewhat arbitrary; for example, \cite{Kanagawa2016} defined it as the location where density reaches half of the unperturbed density, \cite{Dong2017} defined it as the location where density reaches the geometric mean between the disk's and the gap's density, and \cite{Zhang2018} defined it as half the distance between the local gas minimum and maximum. All of these definitions led to different scalings. For our purpose, we want $r_{\rm gap}$ to be located at where density is not significant depleted, but also where eccentricity is high; in this case, eq. \ref{eq:rgap} is the more optimal choice.

\begin{figure}[h!]
    \centering
    \includegraphics[width=0.95\linewidth]{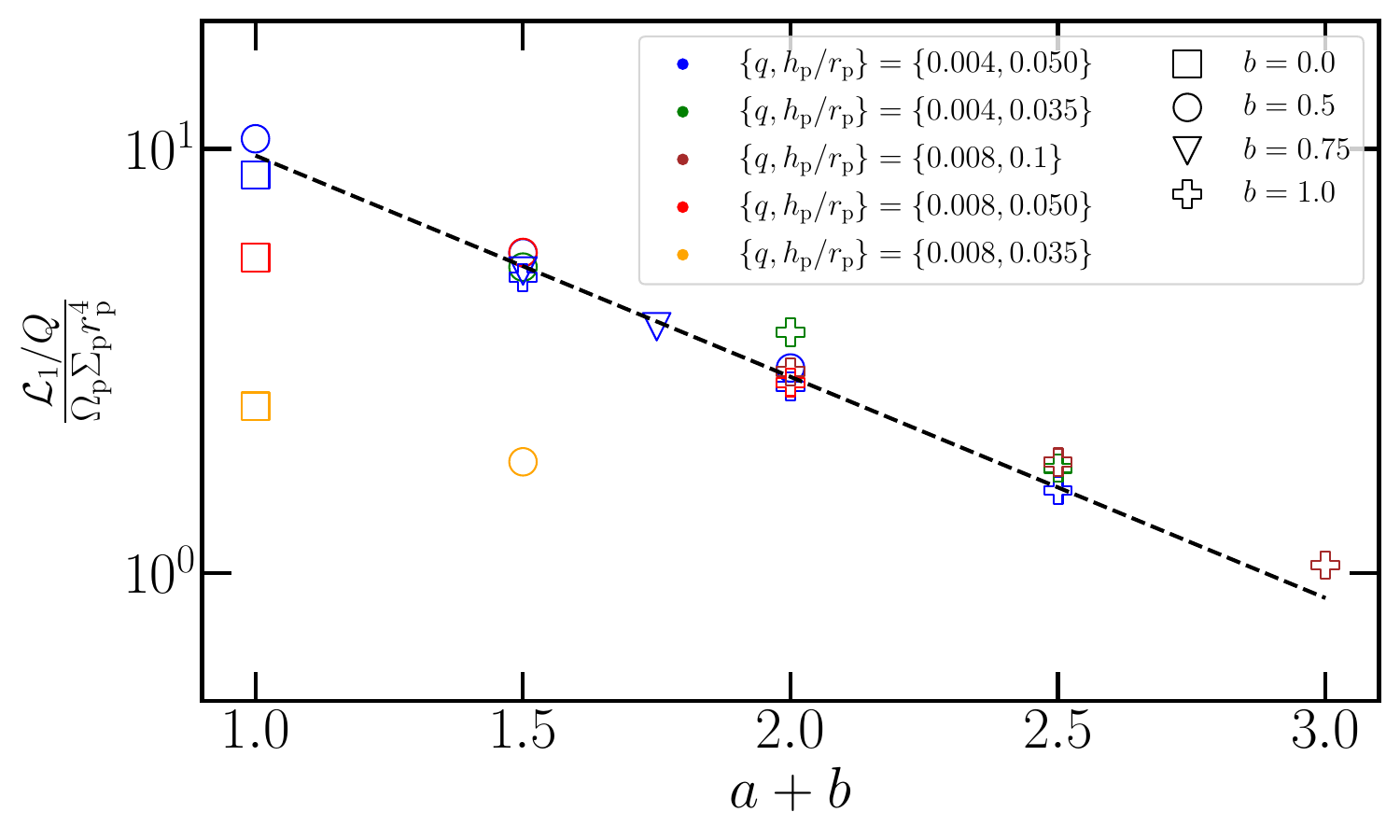}
    \caption{$\mathcal{L}_1$, normalized by $Q$ (eq. \ref{eq:Q}), versus $a+b$. Two strong correlations emerge in this plot: 1) $\mathcal{L}_1$ is mostly proportional to $Q$, the ratio that describes eccentricity excitation vs damping; and 2) $\mathcal{L}_1$ is mostly a function of $a+b$, rather than $a$ and $b$ separately. $\mathcal{L}_1/Q$ appears to follow an exponential scaling with $a+b$ similar to the dashed black line, which is described by eq. \ref{eq:L1_fit}. These scalings enable us to predict disk eccentricity from the values of $Q$ and $a+b$.}
    \label{fig:ecc_trend}
\end{figure}

Fig. \ref{fig:ecc_trend} plots $\mathcal{L}_1/Q$ as a function of $a+b$. We observed a tight correlation, suggesting $Q$ is indeed a predictive scaling factor. Furthermore, $\mathcal{L}_1$ appears to fall exponentially with increasing $a+b$, which can be roughly described as:
\begin{equation}
    \frac{\mathcal{L}_1}{\Omega_{\rm p}\Sigma_{\rm p}r_{\rm p}^4} \approx 32Q \hspace{1 mm} \exp\left[-1.2(a+b)\right] \, .
    \label{eq:L1_fit}
\end{equation}
This function is plotted on fig. \ref{fig:ecc_trend} as the black dashed line. The origin of the exponential scaling with $a+b$ is unclear, but the idea that the important factor is $a+b$, rather than $a$ or $b$ separately, is perhaps not surprising. We have already reasoned that an equilibrium state is established when pressure force counters eccentricity excitation. The factor $a+b$ is the non-dimensional radial pressure gradient of the disk, and so a higher value of $a+b$ means the disk is more pressure supported, which should make it more resistant to eccentricity excitation. Since $\mathcal{L}_1$ is an integral extracted from our simulations, it should have some dependence on the outer boundary location if $\ell_1$ does not fall very rapidly. To understand this, we next inspect the spatial distribution of $\ell_1$.

\begin{figure*}[]
    \centering
    \includegraphics[width=0.95\linewidth]{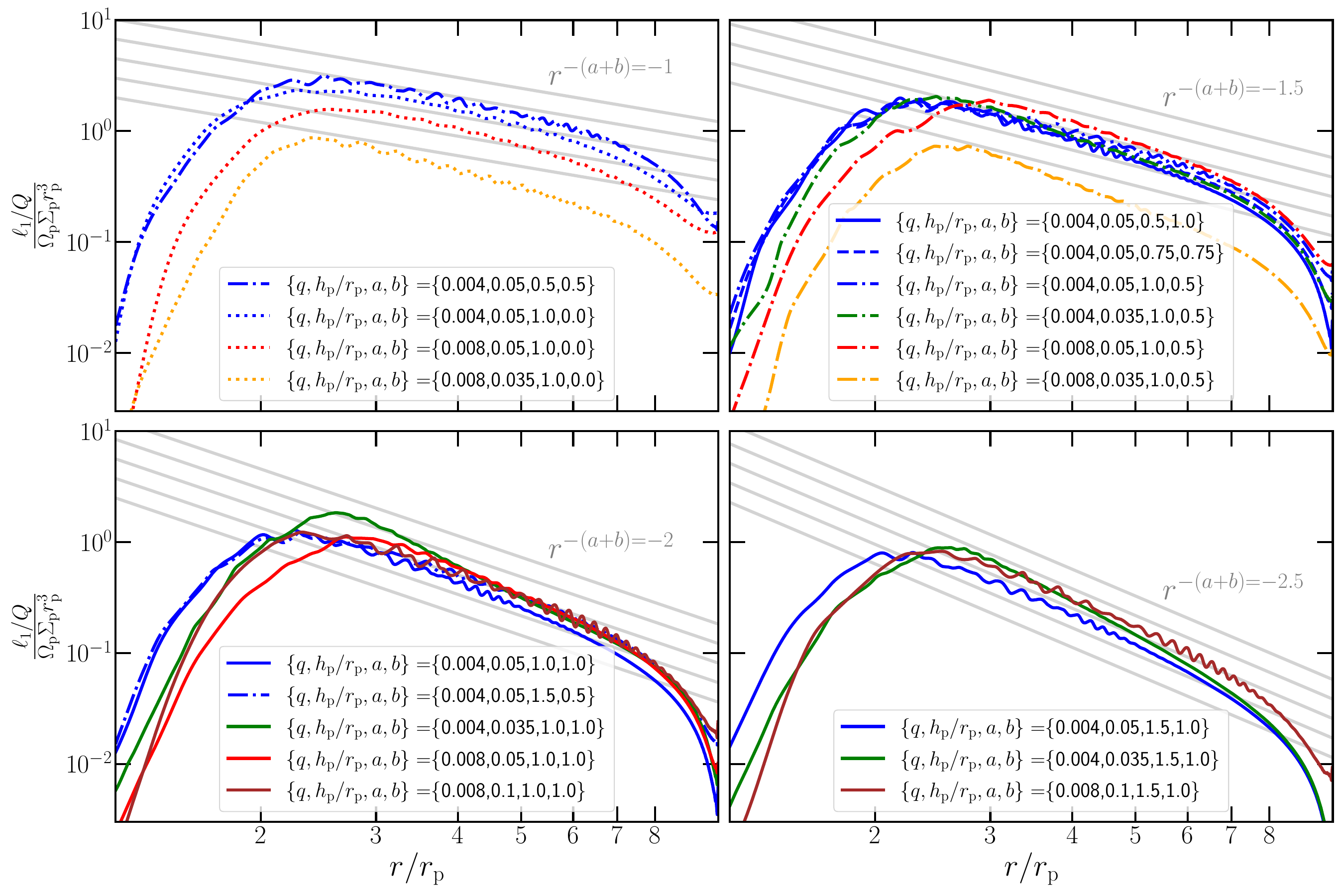}
    \caption{Profiles of $\ell_1$ from different simulations, all normalized by the scaling factor $Q$ (eq. \ref{eq:Q}). The four plots are separated by values of $a+b$: profiles with $a+b=1$ are shown in the upper left, $a+b=1.5$ in the upper right, $a+b=2$ in the lower left, and $a+b=2.5$ in the lower right. Blue lines indicate $\{q,h_{\rm p}/r_{\rm p}\}=\{0.004,0.05\}$; green lines indicate $\{q,h_{\rm p}/r_{\rm p}\}=\{0.004,0.035\}$; brown ones have $\{q,h_{\rm p}/r_{\rm p}\}=\{0.008,0.1\}$; red ones have $\{q,h_{\rm p}/r_{\rm p}\}=\{0.008,0.05\}$, and orange ones have $\{q,h_{\rm p}/r_{\rm p}\}=\{0.008,0.035\}$. Dotted, dot-dashed, dashed, and solid lines correspond to $b=$ 0, 0.5, 0.75, and 1, respectively. The grey lines are power laws that scale with $r^{-(a+b)}$. We find that $\ell_1$ closely follows the $r^{-(a+b)}$ profile in the outer disk, roughly between $2r_{\rm p}$ and $7r_{\rm p}$. Eccentricity is artificially damped near the outer boundary by our circular boundary condition. }
    \label{fig:a_b_profile}
\end{figure*}

The fact that $a+b$ can be used to predict the value of $\mathcal{L}_1 /Q$ also points to the idea that the profile of $\ell_1$ should be determined by $a+b$. In fig. \ref{fig:a_b_profile}, we plot the profiles of $\ell_1$ from different simulations, separated by different values of $a+b$. Overall, we find that $\ell_1$ largely scales as $r^{-(a+b)}$, and only deviates from it either inside the gap, or close to the outer simulation boundary.

Knowing the radial profile of $\ell_1$ allows us to work backwards and approximate the eccentricity profile $e_1$. The bulk of the outer disk remains in steady viscous flow, which means its surface density should be well approximated by our initial, steady-state condition, $\Sigma \propto r^{-a}$. Also, since eccentricity peaks around the gap edge and falls quickly, we can also assume the angular momentum profile to be roughly Keplerian for the bulk of the outer disk. In other words, $\Omega\Sigma r^3$ should scale as $r^{-a+\frac{3}{2}}$. Combining this with our empirical finding that $\ell_1\propto r^{-(a+b)}$, we can infer:
\begin{align}
    e_1 & \approx e_{\rm 1,p} \left(\frac{r}{r_{\rm p}}\right)^{-b-\frac{3}{2}}
    \label{eq:e1_profile}
\end{align}
where $e_{\rm 1,p}$ is some constant of proportionality that may depend on physical parameters such as $q$, $h_{\rm p}/r_{\rm p}$, $a$, and $b$. Next, we aim to evaluate $e_{\rm 1,p}$ and uncover these dependencies.

We integrate our approximate profile of $\ell_1$ and equate it to eq. \ref{eq:L1}. The challenge here is determining the limits of integration. As seen in fig. \ref{fig:a_b_profile}, $\ell_1$ falls off near the planet, i.e., inside the planetary gap. This is due to the combined effects of lower $\Sigma$ and lower eccentricity inside the planet's gap. Through inspection, we determine that $\ell_1$ typically reaches its maximum around $r=2r_{\rm p}$. This is perhaps not surprising since the 1:3 ELR is near $2r_{\rm p}$. Ignoring the small contribution inside the gap, we choose $r=2r_{\rm p}$ to be our inner limit. The outer limit is simply the outer boundary of our simulation, $r=10r_{\rm p}$. We can now approximate $\mathcal{L}_1$ as:
\vspace{-1 mm}
\begin{align}
    \frac{\mathcal{L}_1}{\Omega_{\rm p}\Sigma_{\rm p}r_{\rm p}^4} & \approx  e_{\rm 1,p}  \int_{2r_{\rm p}}^{10r_{\rm p}}\left(\frac{r}{r_{\rm p}}\right)^{-(a+b)}~\frac{{\rm d}r}{r_{\rm p}}\\
    ~& \approx  \frac{e_{\rm 1,p}}{a+b-1} \left(2^{-(a+b)+1} - 10^{-(a+b)+1}\right) \, .
    \label{eq:L1_integral}
\end{align}
Combining this with eq. \ref{eq:L1_fit}, $e_{\rm 1,p}$ can be expressed as:
\begin{equation}
    e_{\rm 1,p} \approx 32Q\frac{a+b-1}{2^{-(a+b)+1} - 10^{-(a+b)+1}} \hspace{1mm} \exp\left[-1.2(a+b)\right] \, .
    \label{eq:e1_appr1}
\end{equation}

\begin{figure}[h!]
    \centering
    \includegraphics[width=0.95\linewidth]{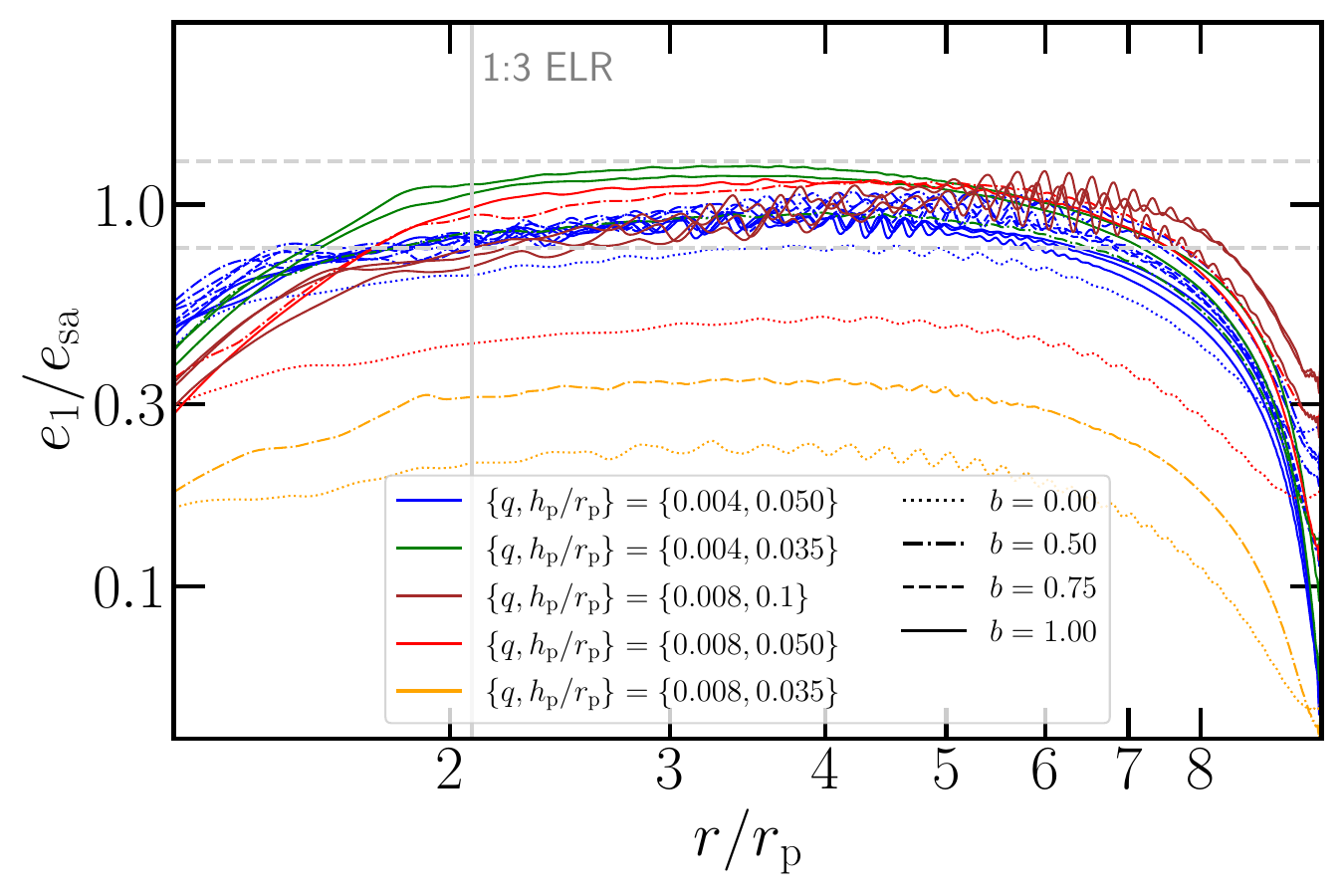}
    \caption{Plots of disk eccentricity $e_1$ divided by our semi-analytic expression $e_{\rm sa}$ (eq. \ref{eq:e1_appr2}). The $a$ values for these curves vary between $0.5$ and $2.0$. The vertical gray line marks the location of the 1:3 ELR, and the horizontal dashed gray lines bracket a $30\%$ difference between $e_1$ and $e_{\rm sa}$, which captures most of our models outward of the 1:3 ELR. We discuss the exceptional case of $\{q,h_{\rm p}/r_{\rm p}\}=\{0.008,0.035\}$ in sec. \ref{sec:wide_gap}.}
    \label{fig:ecc_profile}
\end{figure}

It is worth noting that the location of the outer boundary, $10 r_{\rm p}$, factors in. Clearly, since we are integrating a power law in eq. \ref{eq:L1_integral}, the shallower the power is, the more the integral will lean on the outer limit. When $a+b\gtrsim 2$, setting the outer limit at $10 r_{\rm p}$ changes the integral by less than $20\%$ compared to an outer limit at infinity, but they quickly diverge when $a+b$ approaches 1. Despite this, intriguingly, the value of the right hand-side of eq. \ref{eq:e1_appr1} is narrowly bound even as we vary $a+b$ between $1$ and $3$---it only goes from $6Q$ to $7.5Q$. This suggests that even though eq. \ref{eq:L1_integral} is affected by the outer boundary location, eq. \ref{eq:L1_fit} is affected in nearly the same way and mostly compensates for it. In other words, our empirical relation for $\mathcal{L}_1$ (eq. \ref{eq:L1_fit}) is simply an approximation for the integral in eq. \ref{eq:L1_integral}, indicating the true value for $e_{\rm 1,p}$ should be insensitive to the outer boundary location. For this reason, we take a leap of faith and eliminate the $a+b$ dependency by approximating eq. \ref{eq:e1_appr1} as $e_{\rm 1,p}\approx7Q$. We now arrive at a semi-analytic profile for disk eccentricity, which we denote $e_{\rm sa}$:
\begin{align}
    e_{\rm sa} &= 7 Q \left(\frac{r}{r_{\rm p}}\right)^{-b-\frac{3}{2}}\\
    ~           &= 7 \left(2.08\right)^{-a} \frac{q}{(h_{\rm p}/r_{\rm p})} \left(\frac{r_{\rm gap}}{r_{\rm p}}\right)^{a-\frac{b}{2}-2} \left(\frac{r}{r_{\rm gap}}\right)^{-b-\frac{3}{2}} \,,
    \label{eq:e1_appr2}
\end{align}
where we have expressed radius in reference to either $r_{\rm p}$ or $r_{\rm gap}$; $r_{\rm gap}$ is arguably a better reference point since $r_{\rm p}$ is inside the gap and our expression should not work there. Fig. \ref{fig:ecc_profile} plots $e_1/e_{\rm sa}$ for all of our simulations. We find that $e_{\rm sa}$ is within $30\%$ of $e_1$ for most of the disks, except within the 1:3 ELR or near the outer simulation boundary. We have verified that using the more complicated expression, eq. \ref{eq:e1_appr1}, does not improve the match significantly.

\begin{figure}[h!]
    \centering
    \includegraphics[width=0.95\linewidth]{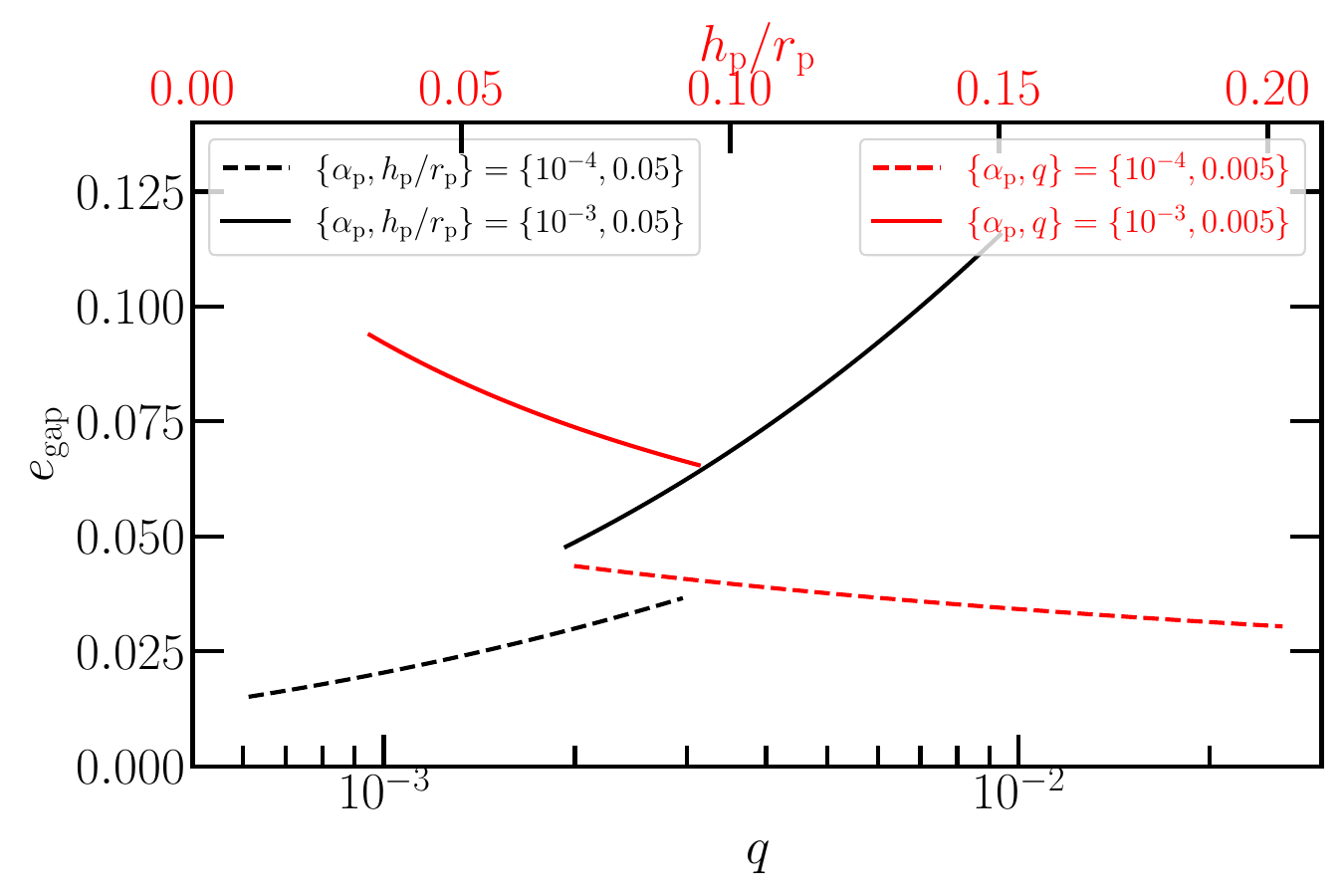}
    \caption{$e_{\rm gap}$ as functions of $q$ and ${h_{\rm p}}/{r_{\rm p}}$ (eq. \ref{eq:e_gap}) with different values of $\alpha_{\rm p}$. The black lines fix ${h_{\rm p}}/{r_{\rm p}}=0.05$ and varies $q$; the red lines fix $q=0.005$ and varies ${h_{\rm p}}/{r_{\rm p}}$; and all lines fix $a=b=1$. The range of values for both $q$ and ${h_{\rm p}}/{r_{\rm p}}$ corresponds to $30<K'<700$, or $2 \lesssim r_{\rm gap} / r_{\rm p}\lesssim 3 $, which ensures the planetary gap is deep enough to promote eccentricity excitation \citep{Tanaka2022}, but not wide enough to deplete the 1:3 ELR (sec. \ref{sec:wide_gap}). Somewhat counter-intuitively, gap edges appear more eccentric when viscosity is higher. This is mainly due to a reduced gap width.}
    \label{fig:e_gap}
\end{figure}

Even though $\ell_1$ may peak near the 1:3 ELR, the gas density there can already be quite low when massive planets open wide gaps. In other words, $r_{\rm gap}$ often exceeds $r_{1:3}$. In practice, observations of eccentric disks would most likely pick up the eccentric gap edges. Using eq. \ref{eq:e1_appr2}, we can estimate how eccentric gap edges may appear by setting $r=r_{\rm gap}$. We refer to this eccentricity as $e_{\rm gap}$:
\begin{equation}
    e_{\rm gap} = 7 \left(2.08\right)^{-a} \frac{q}{(h_{\rm p}/r_{\rm p})} \left(\frac{r_{\rm gap}}{r_{\rm p}}\right)^{a-\frac{b}{2}-2} \, .
    \label{eq:e_gap}
\end{equation}
We overlay ellipses with this eccentricity on top of our simulations in fig. \ref{fig:disks} for a visual comparison. Notably, $e_{\rm gap}$ is only weakly dependent on $q$. The explicit $q$ dependence is partially canceled by the $q$ dependence in $r_{\rm gap}$ because while a higher $q$ strengthens the 1:3 ELR, it also widens the gap. Since eccentricity falls with radius (eq. \ref{eq:e1_profile}), the increased eccentricity is partially reduced by measuring the eccentricity further away from the planet. The same cancellation applies to the $h/r$ dependence as well. Interestingly, this leaves the viscosity dependence. In the limit where viscosity does not dominate eccentricity excitation or damping, it can only affect the gap width. A higher viscosity makes a narrower gap, which makes the gap edge appear more eccentric.

Fig. \ref{fig:e_gap} plots $e_{\rm gap}$ for varying $q$, with $h_{\rm p}/r_{\rm p}=0.05$, $\alpha_{\rm p}=10^{-3}$, $a=1$, and $b=1$ (black solid line). For those chosen parameters, multi-Jupiter-mass planets excite disk eccentricities between $0.05\sim0.1$. If instead we fix $q=0.005$, varying $h_{\rm p}/r_{\rm p}$ also results in a similar range of eccentricity (red solid line). $e_{\rm gap}$ is far more sensitive to $\alpha_{\rm p}$---if $\alpha_{\rm p}$ is reduced to $10^{-4}$, $e_{\rm gap}$ falls below $0.05$ (dashed lines). It should be noted that none of our simulations have $e_{\rm gap}$ values lower than $\sim0.05$. The model with the smallest value of $e_{\rm gap}$ is $\{q,h_{\rm p}/r_{\rm p}, a, b, \alpha_{\rm p}\}=\{0.008,0.1,1,1,10^{-4}\}$, where $e_{\rm gap}=0.05$, so we have extrapolated in our interpretations of fig. \ref{fig:e_gap}. Meanwhile, the model with the highest value of $e_{\rm gap}$ is $\{q,h_{\rm p}/r_{\rm p}, a, b, \alpha_{\rm p}\}=\{0.008,0.05,1,0.5,10^{-3}\}$, where $e_{\rm gap}=0.14$ (excluding the $\{q,h_{\rm p}/r_{\rm p},\alpha_{\rm p}\}=\{0.008,0.035,10^{-3}\}$ set, which we will discuss in sec. \ref{sec:wide_gap}).

We can compare these values with those observed in disks. In MWC 758 \citep{Dong2018,Kuo2022}, the observed eccentricity is about $0.1$. The location of the observed planet MWC 758 c is outside of the eccentric disk \citep{Wagner2023}. The eccentricity of inner disks is beyond the scope of our analysis, but it is possible that an inner, unseen planet, positioned inside the disk cavity, contributes to the observed eccentricity. Although more detail modeling is needed to determine the values of $a$ and $b$, it is likely that this level of eccentricity requires a slightly larger value of $\alpha_{\rm p}$. If we choose, for example, $\alpha_{\rm p}=10^{-3}$, $h/r = 0.05$, and $a=b=1$, then eq. \ref{eq:e_gap} infers that reaching $e_{\rm gap}=0.1$ would require $q=8\times 10^{-3}$, or about 13 Jupiter masses.

In another disk, CI Tau, the observed disk eccentricity is estimated to be about 0.05 \citep{Kozdon2023}. Comparing to MWC 758, a lower eccentricity strongly suggests a lower $\alpha_{\rm p}$, and $h/r$ should also be lower since this eccentric disk is located at a sub-au radius. If we choose $\alpha_{\rm p}=3\times 10^{-4}$, $h/r = 0.03$, and $a=b=1$, then we find that for $e_{\rm gap}=0.05$, we need $q=0.003$, or about 3 Jupiter masses. This is broadly in line with the proposed planet masses given by \cite{Manick2024} and \cite{Donati2024} at an orbital period of around 24 days. 

The eccentricities found in the disks around HD 142527 and IRS 48 are much higher, 0.3$\sim$0.45 \citep{Garg2021} and 0.27 \citep{Yang2023}, respectively. We have demonstrated that these levels of eccentricity are difficult to attain through planet-disk interaction, unless $\alpha_{\rm p}$ is very high. For example, if we choose $\alpha_{\rm p}=5\times 10^{-2}$, $h/r = 0.05$, and $a=b=1$, then we can get $e_{\rm gap}=0.27$ when $q=0.01$, but with an $\alpha_{\rm p}$ value this high, eccentricity damping by viscosity might kick in and circularize the disk. Indeed, for HD 142527, the stellar companion carving out an eccentric cavity is the more likely explanation, as suggested by \cite{Garg2021}. Alternatively, perhaps the eccentricities are not measured at $r_{\rm gap}$, but somewhere closer to $r_{\rm p}$. Since eccentricity follows a power law profile (eq. \ref{eq:e1_appr2}), if there is enough material left inside a gap to be picked up in observations, the observed eccentricity can be higher than $e_{\rm gap}$ for the same disk. Resolving the radial variations of eccentricity would help decide whether the eccentricities in HD 142527 and IRS 48 are of planetary origin.

%%%%%%%%%%%%%%%%%%%%%%%%%%%%%%%%%%%%%%%%%%%%%%%%%%%%%%%%%%%%%%%%%%%%%%%%%%%%%%%%%%%%%%%%%%%
\subsection{When the Gap is Too Wide}
\label{sec:wide_gap}
The $\{q,h_{\rm p}/r_{\rm p},\alpha_{\rm p}\}=\{0.008,0.035,10^{-3}\}$ set produces about three times lower eccentricity than the overall trend we observe (fig. \ref{fig:ecc_trend}). This set also has the highest value of $K'$ (and therefore $r_{\rm gap}$), which suggests that when a gap becomes too wide, eccentricity excitation can weaken. This is expected behavior because, just like how the depletion of the ECRs enables the 1:3 ELR, the 1:3 ELR should also become depleted as the planetary gap becomes wider and wider. The only difference here is that the 1:3 ELR lies further away from the planet, so it takes a wider gap to deplete it. Evidently, our $\{q,h_{\rm p}/r_{\rm p},\alpha_{\rm p}\}=\{0.008,0.035,10^{-3}\}$ set reaches this limit.

\begin{figure}[h!]
    \centering
    \includegraphics[width=0.95\linewidth]{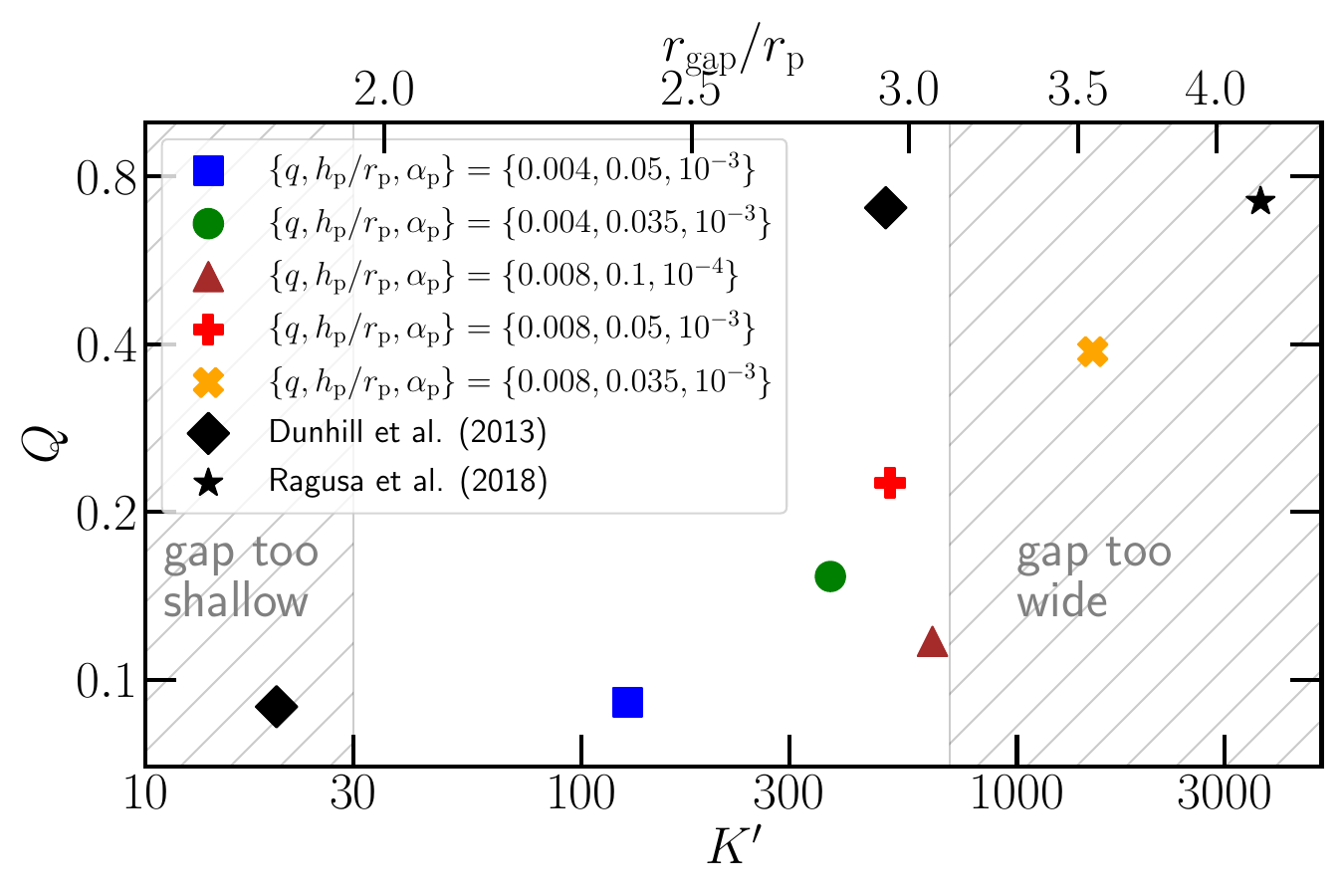}
    \caption{This figure illustrates where our simulations sit in the parameter space when $a=b=1$. When $K'\lesssim30$ (or, equivalently, $r_{\rm gap}\lesssim 2 r_{\rm p}$), \cite{Tanaka2022} found that the planetary gap is too shallow to trigger eccentricity excitation. For comparison, also shown here are the simulations from \cite{Dunhill2013} and \cite{Ragusa2018}. We find that the eccentricity in our $\{q,h_{\rm p}/r_{\rm p}\}=\{0.008,0.035\}$ set is about 3 times lower than expected, which is likely because the gap becomes wide enough to start depleting the 1:3 ELR. This brackets the region of high eccentricity roughly between $30\lesssim K'\lesssim 700$.}
    \label{fig:r_gap}
\end{figure}

Fig. \ref{fig:r_gap} plots the positions of our simulation sets in the $K'$ (eq. \ref{eq:K_prime}) vs $Q$ (eq. \ref{eq:Q}) parameter space when $a=b=1$; the corresponding $r_{\rm gap}$ values are also shown. The scaling relations derived in this paper applies when the 1:3 ELR is both undepleted and unhindered by the ECRs. This requires the planetary gap to be not too shallow, but also not too wide, thus bracketing the parameter space roughly between $K'\gtrsim30$ and $K'\lesssim700$. Also shown on the plot are simulations from \cite{Dunhill2013} and \cite{Ragusa2018}. \cite{Dunhill2013} simulated systems with $\{q,h_{\rm p}/r_{\rm p},\alpha_{\rm p}\}=\{0.005,0.05,10^{-2}\}$ and $\{q,h_{\rm p}/r_{\rm p},\alpha_{\rm p}\}=\{0.025,0.05,10^{-2}\}$, and reported that the disk becomes eccentric only when $q=0.025$. This is consistent with our picture since their $q=0.005$ model has $K'$ less than 30, but their $q=0.025$ model does lie in the sweet spot for eccentricity excitation. \cite{Ragusa2018} simulated a system with $\{q,h_{\rm p}/r_{\rm p},\alpha_{\rm p}\}=\{0.013,0.036,10^{-3}\}$, which lies in the ``gap too wide'' regime; accordingly, they found that disk eccentricity remains low unless the planet migrates quickly. We speculate that when the planet migrates quickly, it is able to move away before fully opening its gap, thus reducing the gap width and keeping the 1:3 ELR alive (see \cite{Rafikov2002} for a description of gap-opening by a migrating planet).

We have so far described a picture where disk goes from circular to eccentric as the planet opens a deeper and deeper gap, and then back to circular when the gap becomes too wide, but numerical work have shown that the disk will again become eccentric when $q\gtrsim0.01$ \citep[e.g.,][]{MacFadyen2008,Miranda2017,Thun2017,Ragusa2020}. In this regime, the ``planet'' should be considered a binary companion, and the ``planetary gap'' becomes a central disk cavity. While the exact origin of eccentricity excitation is unclear in this case, the 1:3 ELR is likely not responsible, since it should be largely depleted. Instead, the quadrupole moment of the binary might be the culprit; for instance, \cite{Munoz2020} demonstrated that a stronger quadrupole moment traps eccentricity closer to the cavity's edge. 

%%%%%%%%%%%%%%%%%%%%%%%%%%%%%%%%%%%%%%%%%%%%%%%%%%%%%%%%%%%%%%%%%%%%%%%%%%%%%%%%%%%%%%%%%%%
%%%%%%%%%%%%%%%%%%%%%%%%%%%%%%%%%%%%%%%%%%%%%%%%%%%%%%%%%%%%%%%%%%%%%%%%%%%%%%%%%%%%%%%%%%%
%%%%%%%%%%%%%%%%%%%%%%%%%%%%%%%%%%%%%%%%%%%%%%%%%%%%%%%%%%%%%%%%%%%%%%%%%%%%%%%%%%%%%%%%%%%
\section{Discussion} \label{sec:Discussion}

It may be useful to compare our results to the evolution of linear eccentric modes, which have been studied extensively \citep[e.g.,][]{Ogilvie2008,Teyssandier2016,Lee2019,Li2021}. These modes are trapped in effective potential wells that typically require pressure gradients that are shaper than a power law to create. The fact that our outer disks, which do follow a power law in pressure, sustain some eccentricity seems to go against this idea. However, what we observe in this work is likely not a trapped, linear eccentric mode---rather, it is a balance of eccentricity excitation and damping. In other words, the 1:3 ELR is constantly working, like a perpetual machine, to restore eccentricity into a disk that would probably circularize otherwise due to the non-linear damping by gas pressure. In reality, this eccentricity excitation should come at the expense of the planet's angular momentum, and it is only perpetual because the planets in our models remain on fixed orbits. This is appropriate in the limit where the disk is much less massive than the planet. If the disk is more massive, one expects the planet's orbit to evolve as a result of its interaction with the disk. For example, the planet's orbit can also become eccentric \citep[][]{Papaloizou2001,Dunhill2013}, and, in the long term, co-evolve with the disk in a way that resembles secular planet–planet interaction \citep{Ragusa2018}. Planetary gas accretion can also have a profound effect; an accreting, eccentric planet can potentially open a gap much wider than those seen in this work. \cite{Muley2019} found that such a planet might end up circularizing the disk even if it was eccentric at an earlier time, simply by eliminating the 1:3 ELR through accretion.

Another limitation in our analysis is that we have focused on the regime where viscosity does not dominate the evolution of eccentricity, which does require that we set the value of $\alpha_{\rm p}$ at or below $10^{-3}$. When viscosity is higher, shear viscosity is able to excite eccentricity if it is implemented using the $\alpha$ prescription, while bulk viscosity damps eccentricity \citep{Lyubarskij1994,Latter2006,Ogilive2001,Latter2006,Lubow2010}. In our setup, the net effect of viscosity should be damping eccentricity (see sec. \ref{sec:Initial_Bound}). Indeed, we have tested and found that when increasing $\alpha_{\rm p}$ to $10^{-2}$, eccentricities in our models damp to negligible values.

In reality, protoplanetary disks are practically inviscid; in numerical models, viscosity functions as a proxy for turbulent diffusion. How turbulence may influence eccentricity evolution, and whether viscosity is a good representation of that, remain topics of active research. For example, turbulent disks subjected to the magneto-rotational instability does not always circularize \citep{Dewberry2020,Chan2022}, and can even change the strength of the 1:3 ELR by redistributing material \citep{Oyang2021}.

Three-dimensional (3D) effects may also play a role in eccentricity evolution. In 3D disks, eccentric motion would cause an imbalance in vertical hydrostatic equilibrium because of the time-varying vertical acceleration, which excites compressional oscillations (``breathing modes'' where periodic vertical expansion/compression occurs along an eccentric orbit; \cite{Ogilvie2008}) . Additionally, 2D gap structure depends on the choice of smoothing length (eq. \ref{eq:potential}), which does not always match their 3D counterparts \citep{Fung2016, Cordwell2025}. This could lead to a change in gas density near the 1:3 ELR and therefore effect the eccentricity growth. These effects could impact our results and should be investigated further.

Finally, our models are locally isothermal. We expect that any polytropic equation of state with an adiabatic index higher than 1 should result in a stronger pressure response when the disk is eccentric, hence increasing the amount of damping. As a result, with all else equal, adiabatic disks should be less eccentric than isothermal ones. Future investigations may look into the effects of non-isothermal equations of state.

\subsection{Disk Precession}
We measure, at each annulus, the argument of periapsis from the complex phase of the $m=1$ mode (eq \ref{eq:e1}). We find that our disks typically remain apsidally aligned in the outer disks (also apsidally aligned in the inner disks, but doing so independently from the outer disk), and precess as a solid body, similar to the eccentric modes described by \cite{Ogilive2019}. Both gas pressure and planetary perturbation can lead to disk precession \citep{Ogilive2001,Goodchild2006,Lubow2010,Ragusa2024}, but not necessarily in the same direction. A detail analysis of the precession rates measured from our simulations will be described in the second paper of this series.

%%%%%%%%%%%%%%%%%%%%%%%%%%%%%%%%%%%%%%%%%%%%%%%%%%%%%%%%%%%%%%%%%%%%%%%%%%%%%%%%%%%%%%%%%%%
%%%%%%%%%%%%%%%%%%%%%%%%%%%%%%%%%%%%%%%%%%%%%%%%%%%%%%%%%%%%%%%%%%%%%%%%%%%%%%%%%%%%%%%%%%%
%%%%%%%%%%%%%%%%%%%%%%%%%%%%%%%%%%%%%%%%%%%%%%%%%%%%%%%%%%%%%%%%%%%%%%%%%%%%%%%%%%%%%%%%%%%
\section{Conclusion} \label{sec:Conclusion}
We quantified the eccentricities of protoplanetary disks when their embedded planets open deep gaps.  We discovered that the steady state eccentricity of such a disk is characterized by the balance between the excitation effect of the 1:3 ELR and the damping effect of gas pressure, assuming viscosity is sufficiently low. The global density and temperature gradients can also play significant roles since the positions of the planet, the 1:3 ELR, and the gap edge all differ. Using the planet-to-star mass ratio $q$, the disk aspect ratio $h/r$, the non-dimensional density gradient $a$, and the non-dimensional temperature gradient $b$, we constructed a semi-analytic description for the disk's eccentricity (eq. \ref{eq:e1_appr2}). We detail our main findings below.

\begin{itemize}
    \item The radial velocity $m=1$ mode amplitude, $e_1$ (eq. \ref{eq:e1}), and its weighted global integral, $\mathcal{L}_1$ (eq. \ref{eq:L1}), are important diagnostics for eccentric gaseous disks. These quantities are similar to eccentricity and angular momentum deficit in orbital dynamics, but are far more informative for our analysis.
    \item The quantity $\mathcal{L}_1$ scales linearly with $q(h/r)^{-1}$ (eq. \ref{eq:L1_fit} and fig. \ref{fig:ecc_trend}), which directly translates to a higher disk eccentricity when the planet's mass is higher and/or the disk is colder.
    \item We derived a semi-analytic expression for the eccentricities near outer gap edges, $e_{\rm gap}$ (eq. \ref{eq:e_gap}), which agrees with numerical results to within 30\% (fig. \ref{fig:disks} and \ref{fig:ecc_profile}).
    \item Eccentricity decreases with radius as $r^{-b-\frac{3}{2}}$ (fig. \ref{fig:ecc_profile}). Consequently, even though a higher planet mass and/or colder disk should boost the disk's eccentricity, it is partially compensated by the gap edge being further away \citep[eq. \ref{eq:rgap}; ][]{Kanagawa2016}, resulting in only moderate changes in eccentricity (eq. \ref{eq:e_gap} and fig. \ref{fig:e_gap}).
    \item The disk will again circularize if the gap becomes too wide and start depleting the 1:3 ELR. We estimate that this happens when $K'\gtrsim700$. Combining with a lower limit previously established by \cite{Tanaka2022}, our descriptions are most applicable when $30\lesssim K'\lesssim700$ (fig. \ref{fig:r_gap}).
\end{itemize}

%%%%%%%%%%%%%%%%%%%%%%%%%%%%%%%%%%%%%%%%%%%%%%%%%%%%%%%%%%%%%%%%%%%%%%%%%%%%%%%%%%%%%%%%%%%
%%%%%%%%%%%%%%%%%%%%%%%%%%%%%%%%%%%%%%%%%%%%%%%%%%%%%%%%%%%%%%%%%%%%%%%%%%%%%%%%%%%%%%%%%%%
%%%%%%%%%%%%%%%%%%%%%%%%%%%%%%%%%%%%%%%%%%%%%%%%%%%%%%%%%%%%%%%%%%%%%%%%%%%%%%%%%%%%%%%%%%%
\begin{acknowledgments}
JF thank Enrico Ragusa for helpful correspondence. This research used in part resources on the Palmetto Cluster at Clemson University under National Science Foundation awards MRI 1228312, II NEW 1405767, MRI 1725573, and MRI 2018069. The views expressed in this article do not necessarily represent the views of NSF or the United States government.
\end{acknowledgments}

%%%%%%%%%%%%%%%%%%%%%%%%%%%%%%%%%%%%%%%%%%%%%%%%%%%%%%%%%%%%%%%%%%%%%%%%%%%%%%%%%%%%%%%%%%%
\software{\texttt{PEnGUIn} \citep{Fung_thesis}, \texttt{NumPy} \citep{numpy}, \texttt{SciPy}
\citep{scipy}, \texttt{Matplotlib} \citep{matplotlib}}

%%%%%%%%%%%%%%%%%%%%%%%%%%%%%%%%%%%%%%%%%%%%%%%%%%%%%%%%%%%%%%%%%%%%%%%%%%%%%%%%%%%%%%%%%%%
\bibliography{references}
\bibliographystyle{aasjournal}

\end{document}